\documentclass[aps,prb,twocolumn,superscriptaddress,showpacs,amsmath,amssymb,longbibliography]{revtex4-1}
\usepackage[english]{babel}
\usepackage{amsmath}
\usepackage{bm}
\usepackage{graphicx,bbm} 
\usepackage{times}
\usepackage{epsfig} 
\usepackage[utf8]{inputenc}
\usepackage[colorlinks,linkcolor=blue,citecolor=blue,urlcolor=blue]{hyperref}
\usepackage{color}
\usepackage{latexsym}
\usepackage{ulem}
\definecolor{nred} {RGB}{224,0,0}
\definecolor{nblue} {RGB}{28,130,185}
\definecolor{dgreen} {RGB}{78,138,21}
\definecolor{norange}{RGB}{230,120,20}

\begin{document} 
\title{Reduced-basis approach to many-body localization}
\author{P. Prelov\v{s}ek}
\affiliation{Jo\v zef Stefan Institute, SI-1000 Ljubljana, Slovenia}
\affiliation{Faculty of Mathematics and Physics, University of Ljubljana, SI-1000 
Ljubljana, Slovenia}
\author{O. S. Bari\v si\'c}
\affiliation{Institute of Physics, Zagreb, Croatia}
\author{M. Mierzejewski}
\affiliation{Department of Theoretical Physics, Faculty of Fundamental Problems of Technology, 
Wroc\l aw University of Science and Technology, 50-370 Wroc\l aw, Poland}

\date{\today}
\begin{abstract}
Within the standard model of many-body localization, i.e., the disordered chain of spinless fermions, we investigate 
how the interaction affects the many-body states in the basis of noninteracting localized Anderson states. From this 
starting point we follow the approach that uses a reduced basis of many-body states. Together with an extrapolation 
to the full basis, it proves to be efficient for the evaluation of the stiffnesses of local observables, which remain finite 
within the non-ergodic regime and represent the hallmark of the many-body localization (MBL).  The method enables a larger span of 
system sizes and, within the  MBL regime, allows for a more careful analysis of the size scaling of calculated quantities. 
On the other hand, the survival stiffness as the representative of  non--local quantities, reveals limitations of the reduced-basis approach.

\end{abstract}

\maketitle


\section{Introduction}

The interplay of disorder and particle interaction (correlations) has proven to be a challenging theoretical problem. While 
the Anderson localization of noninteracting particles is a well established phenomenon in disordered systems,\cite{anderson58,mott68} 
the role of particle interaction is not fully clarified. In particular, whereas the interacting fermionic systems have been addressed quite 
early-on,\cite{fleishman80} the realization of many-body localization (MBL) has been seriously considered only within the 
last decade.\cite{basko06,oganesyan07} By now, numerous theoretical studies, mostly numerical on small systems, have 
established key properties of the MBL systems quite well. Those particularly relevant for the present study are: a) a non-ergodic 
behavior of all correlation functions and the absence of thermalization of quenched initial 
states,\cite{monthus10,pal10,luitz16,mierzejewski16,prelovsek217} and b) the existence of local integrals of motion 
(LIOMs).\cite{serbyn13,huse14,ros15,imbrie16} While these two conditions for the MBL are evidently related, 
it is not trivial to derive the LIOMs and establish explicitly the connection with the nonergodic behavior.\cite{serbyn15,chandran15,
rademaker16,mierzejewski17}

Experimental tests of the MBL physics were up to now restricted to trapped ions,\cite{smith2016} mainly to  
optical lattices of fermionic\cite{schreiber15,bordia16,luschen16,choi16} and bosonic\cite{choi16} cold atoms, with the focus set on mostly the 
abovementioned properties of the MBL. The experiments investigating one-dimensional (1D) or quasi-1D systems, by following the decay 
of imbalance have find that at large disorder the system can become  nonergodic.\cite{schreiber15,bordia16,luschen16} A similar consideration 
of interacting fermions on a three-dimensional (3D) optical lattice have shown the vanishing of the d.c. mobility,\cite{kondov15} 
which is one of the hallmarks of MBL.\cite{berkelbach10,barisic10,agarwal15,barlev15,steinigeweg15,barisic16}

The problem of confirming the existence of non-ergodic (MBL) phase and its properties by numerical studies remains a challenge (for similar reasons as in cold-atom experiments). Due to extremely slow dynamics of MBL systems,\cite{mierzejewski16} in order to reach non-ergodic finite values of correlation functions, one needs control over system time evolution up to very long (decay) times  (or very high frequency resolution). The finite long-time values of correlation functions correspond to finite stiffnesses, being the signature of the MBL phase and are therefore in focus of the present study. Long evolution times (or equivalently low frequencies) put severe limitations on the application of sophisticated numerical finite-temperature dynamical methods (for a review see Ref.~\onlinecite{prelovsek17}) and the exact-diagonalization (ED) is therefore used in most of the studies, significantly limiting accessible system sizes. Yet, finite-size effects cannot be ignored because of a possible nontrivial scaling with system size.\cite{PhysRevX.7.021013, PhysRevLett.119.150602,Ponte_2017} To partly overcome this problem, we propose an approach that uses the relevant many-body (MB) basis. It exploits the properties of the MBL phase and, instead with the full MB space, works with a specially constructed reduced basis (RB). It this way, we can better follow the system size scaling of relevant quantities in the long time limit.

Most theoretical studies concerning MBL physics so far have been restricted to the prototype model of MBL, i.e., the disordered 1D model of interacting spinless fermions. In this paper we argue that for strong disorder a good starting point for investigations is that of non-interacting (NI) fermions,\cite{kramer93} when a finite but weak interaction should retain a localized character of the system,\cite{basko06,ros15,imbrie16,kozarzewski16} exhibiting MBL properties as defined above. In this work, the reduced-basis approach (RBA) to the MBL problem is defined in two steps. In the first one, the solution of the corresponding Anderson NI problem is obtained numerically. Then, in the next step, the interaction is introduced and the convergence of results is considered by increasing the number of MB states taken into account via the interaction term. The RBA, combined with the adequate extrapolation to the full basis of MB states, should be able to study a larger span of finite-size systems. Moreover, the goal of the method is to enable to go beyond the prototype model discussed here and allow for the studies of more complex models (as, e.g., the challenging Hubbard model \cite{prelovsek16}) as well as for higher dimensional MBL physics.

Within the RBA, we concentrate at present on the long-time ($t  \to \infty$) correlations or stiffnesses, which are the hallmark of the MBL phase. For the model considered, most direct evidence for the MBL is obtained from the density stiffnesses,  $C_d$, defined as  the infinite--time correlation functions for the particle--number operators  at sites separated by a  distance $d$. We study both the onsite\cite{luitz16} $d=0$  and the off-site $d>0$ correlations, being related to the imbalance stiffness,\cite{mierzejewski16} as directly measured in cold-atom experiments.\cite{schreiber15,bordia16,luschen16} The convergence of $C_d$ with the system size $L$ is related to the existence of LIOMs $Q_l$,\cite{mierzejewski17} whereby, in the MBL regime, by using the RBA we can study larger systems than before in numerical studies.\cite{mierzejewski16} In this context, it is important to have an ability to perform a scaling of results to the full MB space. We argue that the convenient parameter for this scaling is $\ln{N_{st}}$,\cite{prelovsek93} where $N_{st}$ is the number of MB states taken into account. In addition, we consider the survival stiffness $S$ (or the MB inverse participation ratio),\cite{heyl14,torres15} which is a non--local quantity that characterizes (in contrast to the local stiffness $C_d$) the convergence of the many-body wave function. $S$ is dependent on the system size $L$,\cite{torres15} even within the MBL phase.  Results confirm the exponential dependence of $S$ on $L$ and in this respect reveal the limitations of the RBA for non--local quantities.

In Sec.~II we present the model by rewriting it in the basis of NI localized (Anderson) states. In Sec.~III A we introduce a procedure, resembling to the degenerate perturbation theory, to generate relevant RB starting from an initial MB Anderson product state. The full diagonalization within such restricted MB basis allows for an evaluation of observables, in our case the density and the survival stiffnesses. The extrapolation procedure to the unrestricted result at given $L$ is also presented.  In Sec~III B we analyze the statistical properties of interaction matrix elements, in particular the role of cubic vs. quartic interaction terms. In Sec.~IV we introduce and evaluate the  local-density stiffnesses, which represents the main criteria for the MBL, and discuss their relation to LIOMs.  We show that for large disorders within a given random configuration and at fixed system size $L$, the RBA results for the local stiffnesses can converge fast.  Configuration-averaged results for the onsite $C_0$ and nearest neighbor local stiffness $C_1$ are presented, with extrapolations towards the full basis for given $L$, allowing for the finite-size scaling of $C_0$.  In Sec.~VI we consider the survival stiffness $S$, representing the non--local criterion for the convergence of the MB 
wavefunction within the RBA. It is shown that $S$ depends exponentially on $L$, while the related exponent varies strongly with disorder.
 
\section{Model in the localized basis}
 
We consider the prototype model of MBL, i.e. the 1D model of interacting spinless fermions with random local potentials, 
\begin{eqnarray}
H &=& - t_h \sum_{i} \left( c^\dagger_{i+1} c_i + \mathrm{H.c.}\right)
+ \sum_i \epsilon_i n_i\ + V \sum_i n_{i+1} n_{i} \nonumber \\
&=& H_0 + H_V, \label{tv}
\end{eqnarray}
where the quenched disorder in the NI  part $H_0$ is given by the uniform distribution 
$-W < \epsilon_i <W$. On the chain with $L$ sites with periodic boundary conditions we focus on the half-filled 
case with $N=L/2$ fermions, corresponding to the average site occupation  $\bar n= 1/2$. 
Hereafter, the hopping integral is used as the unit of energy, $t_h=1$. 
Regarding the interaction, we study the $V=1$ case, which is halfway between the NI fermions and the isotropic 
Heisenberg point value of $V=2$. With $V=1$, the value of $W_c=4$ has been estimated \cite{lev15} 
as the location of the MBL transition (or crossover).

Our approach starts by solving the $H_0$ part first. In the next step, the structure of the $H_V$ part of the problem is analyzed 
in the MB basis, in which $H_0$ is diagonal, in order to distinguish between resonant (perturbatively large) and 
nonresonant (perturbatively small) contributions. In particular, $H_0$ in Eq.~(\ref{tv}) corresponds to the Anderson problem, 
which can easily be diagonalized numerically for large systems,
\begin{equation}
H_0 = \sum_l \epsilon_l  \varphi^\dagger_l \varphi_l, \qquad 
\varphi^\dagger_l   = \sum_i \phi_{li} c_i^\dagger .\;\label{asp01}
\end{equation}
The single-particle states in Eq.~(\ref{asp01}) may be used to define the basis of localized MB states, 
\begin{equation}
|n \rangle = \prod_l (\varphi^\dagger_l)^{n_l} | 0 \rangle, \qquad  
E^0_{n}= \sum_l n_l \epsilon_l\;,\label{tv3}
\end{equation}
mixed by the interacting part $H_V$,
\begin{eqnarray}
H_V &=& V  \sum_{jklm} \chi_{jk}^{lm} ~\varphi^\dagger_l \varphi^\dagger_m \varphi_k \varphi_j, \nonumber \\
\chi_{jk}^{lm} &=& \sum_i  \phi^*_{li} \phi^*_{m,i+1} \phi_{k,i+1} \phi_{j,i} , \label{hloc}
\end{eqnarray}
where the matrix elements satisfy
\begin{equation}
\chi_{jk}^{lm}=(\chi_{lm}^{jk})^*=-\chi_{kj}^{lm}=-\chi_{jk}^{ml},
\end{equation}
according to fermion anticommutation relations.

Using the $|n \rangle$ representation given by Eq.~(\ref{tv3}), three different sorts of contributions to $H_V$ may be 
distinguished. The first is diagonal, representing the Hartree-Fock correction $H_d$,
\begin{equation}
H_{d}= 2 \sum_{k>l} ( \chi_{lk}^{lk} - \chi_{lk}^{kl} ) n_l n_k\;, \label{hf}
\end{equation} 
which modifies unperturbed eigenenergies, 
\begin{equation}
\tilde E^0_n=E^0_n + \langle n|H_d | n\rangle . \label{ldef} 
\end{equation}
With respect as how they act on the unperturbed $|n\rangle$ state, the two remaining sorts 
of contributions, the cubic $H'_3$ and the quartic $H'_4$, may be interpreted as a single and a double 
creation of electron-hole pair, respectively, 
\begin{eqnarray}
H'_{3}&=& 2 V \sum_{jkm} h^{3}_{jkm}, \;  \\
H'_{4}&=&V\sum_{\substack{j>k \\ m>l}} h^{4}_{jkml},  \\
 h^{3}_{jkm}&=& ( \chi_{jk}^{jm} - \chi_{jk}^{mj} ) n_j \varphi^\dagger_m \varphi_k\;, \label{labh3p}\\
h^{4}_{jkml} &=&
(\chi_{jk}^{lm} + \chi_{kj}^{ml}- \chi_{kj}^{lm} - \chi_{jk}^{ml})
\varphi^\dagger_l \varphi^\dagger_m \varphi_k \varphi_j\;,  \label{labh4p}
\end{eqnarray}
where for both, $H'_3$ and $H'_4$, indices should be different, $j\neq k \neq l \neq m$.

\section{Reduced-basis approach}

Our primary goal is to explore properties of MBL regime and introduce the method with the potential
to extend the capabilities of present numerical approaches, both qualitatively and quantitatively. In this context, the basis consisting of 
MB localized states  $|n \rangle$,  Eq.~(\ref{tv3}), is a natural choice to start considering the effects of the interaction $H^\prime$. 
The latter is responsible for a delocalization of MB eigenfunction, both in the ergodic phase and 
in the MBL regime. Still, the latter effect is more restricted within the MBL phase, allowing, at fixed $L$, 
the efficient choice of RB provided that $V$ is weak enough. 
Since the MB states form continuum spectra (in the $L \to \infty$ limit), any perturbative expansion 
in $H'$ faces the problem of resonances, involving MB states with almost degenerated energies and 
nonzero matrix elements of $H'$ between them. The latter problem is clearly related to challenges in the analytical
considerations of the existence of the MBL.\cite{basko06,ros15,imbrie16}
It has also the central role in our numerical RBA in which the perturbatively large  contributions (resonances) are taken 
into account exactly (to infinite order in $H'$), while small perturbations are neglected.  

\subsection{Reduced-basis construction}

The current approach adopts a strategy that has many analogies with degenerate perturbation theory. 
At fixed $L$, the reduced-basis of MB states $|n \rangle$ is constructed iteratively through a number of generations $G$. 
A new generation is obtained by acting with all terms $h^{3}$ and $h^{4}$  (Eqs. (\ref{labh3p}) and  (\ref{labh4p}), respectively)
on all the states of a previous generation $G-1$,  keeping only those new states that satisfy the resonant condition,
\begin{equation}
\Big|\frac {\langle n^{G-1}| H' |m^G\rangle}{\tilde E^0_0 - \tilde E^G_m } \Big|=r > R\;,\label{R1}
\end{equation}
where $R$ is a resonance threshold parameter that is used to control the convergence of results 
and   $\tilde E^G_m$ is defined via Eq. (\ref{ldef}).
In the spirit of the Rayleigh-Schr\"odinger perturbation-theory series for the eigenstate energy, our formulation of the 
resonance condition, Eq.~(\ref{R1}), involves the  energy $\tilde E^0_0$ of the initial $G=0$ state $|n \rangle$. 

The described procedure for chosen resonance parameter $R$ and the number of generations $G$
defines the creation of RB for given initial  MB state $| n \rangle $.
The complete RB is then  $| \underline n \rangle = \{ | n \rangle,
| n^1 \rangle \cdots | n^G \rangle \}$. The Hamiltonian $H'$, which is only offdiagonal 
within this basis, has predominantly the block structure. 
After creating the RB set, the final step is the (full) exact diagonalization (ED) 
of the Hamiltonian $H$ within this subspace,
\begin{equation}
H |\tilde m \rangle= E_ {\tilde m} | \tilde  m \rangle,  \qquad   | \tilde m \rangle = 
\sum_{\underline n}^{N_{st}} \alpha_{\tilde m, \underline n} | \underline n \rangle, \label{hprime} 
\end{equation}
where $|\tilde m \rangle$ are the eigenstates within RB, and 
$E_{\tilde m}$  the corresponding eigenstate energies. Within the determined
RB we can also evaluate any relevant observable, in particular correlation functions and stiffnesses considered herein.  

\begin{figure}[tb]
\includegraphics[width=0.8\columnwidth]{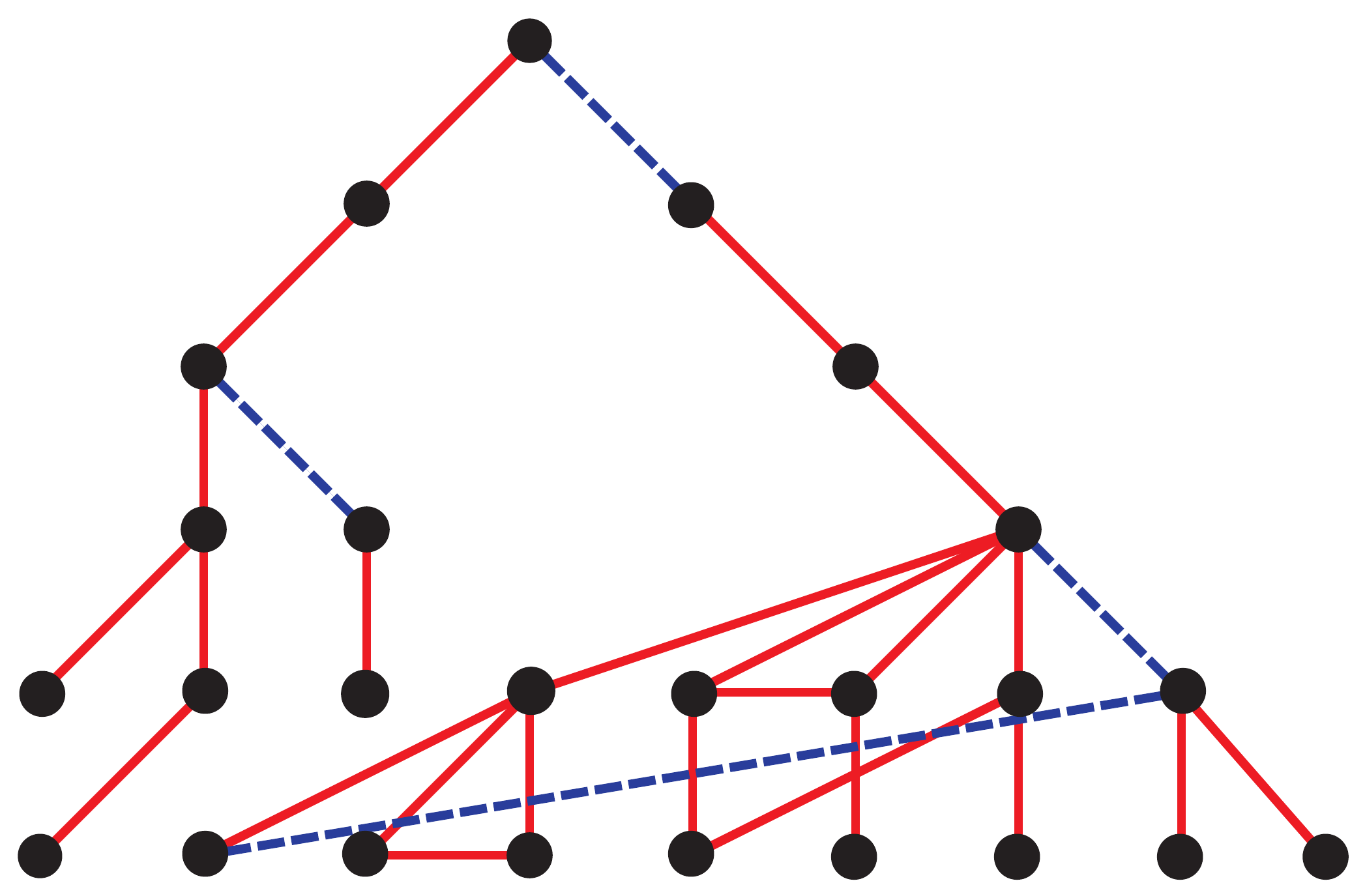}
\caption{(Color online) Connectivity scheme of the reduced basis approach in the MBL regime. The full (red) and
the dashed (blue) lines represent the contributions of the $H_3^\prime$ and the $H_4^\prime$ term, respectively.}
\label{fig1}
\end{figure}

The RBA is at this stage restricted by the requirement imposed by the ED of the RB Hamiltonian matrix. 
The latter may be numerically achieved for $N^{max}_{st} \sim 20000$ states. At half-filling, the full Hilbert space involves 
$N_{tot} = L!/[(L/2)!]^2$ states, limiting standard ED approaches to sizes $L\leq 16$.  
The same is true for RBA with $R=0$. However, by focusing on perturbatively most significant contributions (resonances) 
with $R>0$  and finite $G$, RBA can substantially increase available $L$.  Furthermore, in the MBL regime, the RBA reveals a 
connectivity scheme for resonances, shown in Fig.~\ref{fig1}, which is substantially modified with respect to 
the $R=0$ structure, although it generally does not have a Caylee-tree structure.
At half-filling, the $R=0$ case would involve $\propto L^2$ and $\propto  L^4$ connections 
per one MB state for the $H'_3$ and the $H'_4$ part of the interaction, respectively. On contrary, as schematically shown in Fig.~\ref{fig1}, 
the RBA connectivity is much more sparse, being strongly disorder and state dependent. By changing $R$ in Eq.~(\ref{R1}) we  
control the RB increase per generation. Small $R$ means that the maximal 
number of allowed states $N_{st}^{max}$ is reach early in $G$. Large $G$, on the other hand, means that 
resonances involving high-order contributions in $H'_3$ and $H'_4$ are taken into account. 
The implementation of the RBA therefore represents a tradeoff between small $R$ and large $G$. 

For static quantities the convergence of the results may be analyzed in terms of 
the effective number of states, $N_{st}$, taken into account.
In the following we present the argument for the observation that (at given $L$) the convenient 
parameter for the extrapolation to full Hilbert space $N_{tot}$  is \cite{prelovsek93} 
\begin{equation}
 x = \ln(N_{tot}/N_{st})\;. \label{x}
\end{equation}
Let us assume that $v=V/W <1 $ is the relevant (small) parameter for the perturbation expansion
of quantity $\Gamma$ at fixed $L$,  i.e., $\Gamma=\Gamma^{(0)}+ \Gamma^{(1)}+\cdots$, 
$\Gamma^{(p)} \propto |v|^p$ .  With $M$ being the connectivity of states, the $p$-th order of the perturbative 
expansion involves $N_{st} \propto M^p$ states, so that
$p =\ln{N_{st}}/\ln{M}$. The error of truncating the perturbative series at this order (assuming $p>0$) is then 
$\Delta\Gamma^{(p)} \propto |v|^{p}= N_{st}^{-r}$ where $r =- \ln{|v|}/\ln{M}$. 
Since $M \gg 1/|v|$ one obtains $r \ll 1$ for the current problem of the MBL. In this limit, the power law 
approaches the logarithmic dependence given by Eq.~(\ref{x}), taking into account the limiting 
condition that $x=0$ for $N_{st}=N_{tot}$  as well.\cite{prelovsek93} 
The scaling of results with parameter $x$ in Eq.~(\ref{x}) is particularly convenient since it is independent of $L$, 
permitting a direct comparison of results for systems of different sizes $L$. 

Still, the straightforward  linear  extrapolation towards $x \rightarrow 0$ may not apply. 
In particular, within the MBL regime the results could saturate already for certain $x_0 >0$ (and corresponding $N^0_{st}$) simply due to non--ergodic dynamics. 
The exact MB wave functions in the MBL regime have nonzero projections only  on certain subsets of localized states which 
are much smaller than the dimension of the Hilbert space.  Then, $N^0_{st}$  roughly estimates the average number of $|n \rangle $ 
states with non--vanishing projections on individual MB eigenstates. 

\subsection{Interaction matrix elements} 

A relevant question arises, namely which interaction terms, $H'_3$ or $H'_4$, are more 
important for the construction of the RB. In the case of weak localization,\cite{basko06,imbrie16,ros15} it is 
argued that the relevant terms are the quartic ones $H^\prime_4$, i.e., the double creation of the particle-hole pairs. 
Here, we consider the case of moderate to strong disorder ($W>2$), where the averaged single-particle localization 
length is comparable to the lattice constant, i.e.,  $\xi \sim O(1)$. With the short-ranged localization length $\xi$, 
matrix elements $\chi_{jk}^{lm}$, given in the Anderson basis by Eq.~(\ref{hloc}), are short-ranged as well.  
We can make this statement more objective, if we order the Anderson states not by energy,
but by  their maximum amplitude on the lattice. In the case of $\xi \sim O(1)$ such a
labeling is meaningful and feasible. From Eq.~(\ref{hloc}) it is evident that $\chi_{jk}^{lm}$ in this case
decay fast with increasing distance between real-space sites $j,k,l,m$. On the other hand,
there are restrictions on the denominator. In 1D, one can realize that
Anderson single-particle levels cannot be close in energy, when sites are close in space.
A simple analysis \cite{mott68} provides the constraint for nearest-neighbor sites 
$|\epsilon_{j+1} - \epsilon_j| > 2t_h$, and similarly for next-nearest neighbors  
$|\epsilon_{j+2} - \epsilon_j| > 2t_h^2/W$ etc. Yet, a rigorous analytical analysis of relevant matrix elements as a function of $W$ is complicated. Therefore, we rely on the 
numerical evaluation of their statistics.
 
\begin{figure}[tb]
\includegraphics[width=1.0\columnwidth]{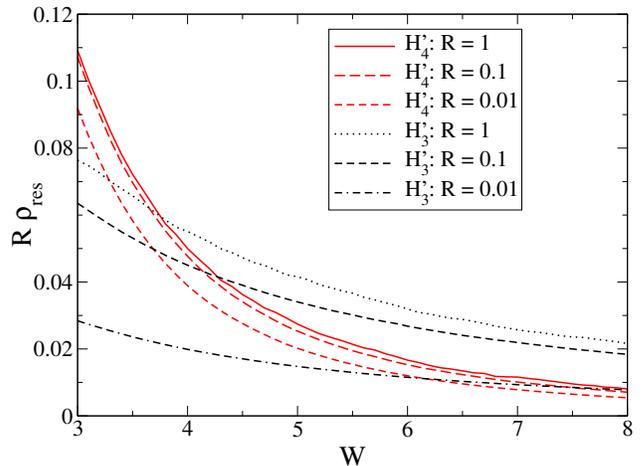}
\caption{(Color online) Density of resonances per one RB state rescaled by $R$, $R\varrho_{res}$, for $V=1$. The results for the $H'_3$ and the $H'_4$ 
part of the interaction are given by black and red curves, respectively. For $W\gtrsim3$, $\varrho_{res}$ scales approximately as $1/R$, 
exhibiting no significant $L$-dependence for $R\gtrsim0.01$.}
\label{fig2}
\end{figure}

In Fig.~\ref{fig2}, we present the numerical results for the average density of resonances $\varrho_{res}$ per one RB state for $V=1$, 
rescaled by $R$, as a function of disorder $W$,
\begin{equation}
R\varrho_{res}=\frac{R}{L}\int_R^\infty dr\;f(r; W)\;,\label{resdist}
\end{equation}
where $f(r; W)$ is the distribution function of resonances according to their strength, $r$, defined in Eq.~(\ref{R1}). In Fig.~\ref{fig2}, the density of resonances is plotted separately for the $H'_3$ and the $H'_4$ term of the interaction. As expected, on the ergodic side of Fig.~\ref{fig2}, $W\sim3$, the density $\varrho_{res}$ rapidly increases as $W$ is decreased. In the opposite direction, for $W>3$, $\varrho_{res}$ becomes small, meaning that the strong resonances satisfying the condition in Eq.~(\ref{R1}) are rare combinations of matrix elements and energy differences. For these rare strong resonances, from the scaling behavior in $R$ shown in Fig.~\ref{fig2}, $\varrho_{res}\sim 1/R$, it is apparent that $\varrho_{res}$ is given by the long tail of the distribution $f(r; W)$, behaving asymptotically as $f(r; W)\propto r^{-\zeta}$, $\zeta\approx2$. Deviations from this simple scaling behavior develop for the $H'_3$ resonances in Fig.~\ref{fig2} as $R$ approaches small values, $R\sim0.01$, whereas even for these small values, $R\varrho_{res}$ remains almost constant for the $H'_4$ resonances. As it may be noted from Fig.~\ref{fig2}, in the large $W$ limit, the density of resonances is always larger for the $H'_3$ than for the $H'_4$ term. Interestingly, the omission of the Hartree-Fock term (\ref{hf}) in the definition of the single-electron energies $\tilde E^0_n$ does not change the statistics of resonances shown in Fig.~\ref{fig2},  either qualitatively or quantitatively.
 
\section{Density correlations}

In the presence of the interaction $V>0$, the MB Anderson states $|n\rangle$ are not the eigenstates of the system. Nevertheless, 
these states may be used as quenched initial states or as a convenient choice of the MB basis for deriving thermodynamic averages.

For confirming MBL behaviors, one of the highly relevant quantities is the local stiffnesses, characterizing long-time local correlations. 
Let us consider the time-correlation function given by   

\begin{equation}
\langle A (t'+t) B(t') \rangle =\frac{1}{N_{tot}} \mathrm{Tr} [A (t'+t) B(t')]\;, 
\end{equation}
where $A$ and $B$ are time-dependent local operators and $\ \langle \cdot \rangle $ is the thermodynamic average at given temperature $T$. 
In our study, we concentrate on the $T \to \infty$ limit, when the average for a given random configuration of the disorder involves all the basis states 
$|n \rangle$ with the same probability. In the context of the RBA, the full averaging is replaced by a sampling over $N_c$ states  $|n \rangle$ of the RB. 
Since $|n\rangle$ are not eigenstates of the Hamiltonian, when $N_c < N_{tot}$ the resulting correlation function,
\begin{equation}
\langle A (t'+t) B(t') \rangle \simeq \frac{1}{N_c}\sum_{n=1}^{N_c} \langle  n | A (t'+t) B(t') |n \rangle\;, 
\end{equation}
depends on $t$ and $t'$. Indeed, the particular value of $t'$ is important when one considers the time evolution of the system prepared in an 
initial quenched state $| n \rangle$ as, e.g., in optical-lattice experiments. However, herein we are interested in thermodynamic averages in the $T \to \infty$ limit, 
investigating the time--averaged correlation function given by
\begin{equation}
C_{AB}(t) = \mathrm{lim}_{\tau \to \infty} \frac{1}{\tau} \int_0^\tau  dt'  \frac{1}{N_c}
\sum_{n=1}^{N_c} \langle  n | A (t'+t) B(t') |n \rangle\;.\label{eq15}
\end{equation}
In particular, the stiffness $ D_{AB}$ corresponds to the long-time limit $t\rightarrow\infty$ of $C_{AB}(t)$. It can be obtained via averaging
  Eq.~(\ref{eq15}) over time $t$,
\begin{eqnarray} 
D_{AB}  &=& \mathrm{lim}_{\tau \to \infty} \frac{1}{\tau} \int_0^\tau  dt \; C_{AB}(t) \nonumber \\
&=& \frac{1}{N_c} \sum_{n, \tilde m}  |\langle n | \tilde  m \rangle|^2 A_{\tilde  m \tilde m} B_{\tilde  m \tilde m}\;\label{cabst},
\end{eqnarray}
where a nondegenerate energy spectrum is assumed.

\subsection{Density stiffnesses}

In the limit $L \to \infty$ and for site--averaged operators  $A$ and $B$  in Eq.~(\ref{cabst}),  the  stiffness $ D_{AB}$ should be independent  
of the particular realization of disorder. This is the case,  e.g., for  $A=1/\sqrt{L} \sum_i A_i$ and $B=1/\sqrt{L} \sum_i B_i$,  where $A_i$ 
and $B_i$ are supported in the surroundings of site $i$. However, if one omits the averaging over lattice sites taking  $A=A_j$ and $B=B_l$ 
then the stiffness $D_{AB}$ may depend explicitly on the disorder configuration $\epsilon_i$.
That is, provided that operators are traceless, $\langle A \rangle = 0$ or $\langle  B \rangle  =0$, and satisfy 
$\langle A  H \rangle =0$ or  $\langle B H \rangle =0$, the nonvanishing stiffnesses in Eq.~(\ref{cabst}), $D_{AB} \neq 0$, 
indicates a nonergodic behavior. In other words, $D_{AB} \neq 0$ may be taken as a criterion for the existence of the MBL
(strictly speaking, for the existence of  a LIOM orthogonal to $H$).

With the substitution $A = \delta n_i = n_i/\bar n - 1 $, $B=  \delta n_j = n_j/\bar n - 1$ in Eq.~(\ref{cabst}), we now turn our attention to the local density stiffnesses,
\begin{eqnarray}
C_{i,i+d}&=&  \frac{1}{N_c} \sum_{n, \tilde m}  |\langle n | \tilde  m \rangle|^2
(\delta n_i)_{\tilde  m \tilde m} (\delta n_{i+d})_{\tilde  m \tilde m}\;.\label{cr}
\end{eqnarray}
In the ergodic limit, the stiffness in Eq.~(\ref{cr}) vanishes since the mean charge density at each lattice site is the same, $\bar n$. 
On the other hand, when the eigenstates of the system involve charges fully localized at lattice sites, $n _i=0,1$, 
the stiffness for half-filling takes the maximal value $C_{i,i+d}=1$.
With $d=0$, it should be noted that $C_{ii}$ is equivalent to the well-known Edwards-Anderson order parameter.\cite{edwards75,luitz116}  
In random spin systems this parameter becomes finite within  the spin-glass phase.\cite{edwards75}  
Another parameter related to the stiffness in Eq.~(\ref{cr}), measured in cold-atom systems,\cite{schreiber15,bordia16,luschen16} is the imbalance $I$
for occupations of two sublattices. The latter represents the stiffness of the staggered 
density operator $n_{q=\pi}=(-1)^{i} n_i$. In terms of $C_{i,i+d}$, it may be expressed as
 \begin{eqnarray}
I  &= &  \frac{1}{L} \sum_d (-1)^{d}\sum_i C_{i,i+d}  \nonumber \\ 
 & = & \frac{1}{\bar{n}^2 N_c L} \sum_{n, \tilde m}  |\langle n | \tilde  m \rangle|^2    \left[  \sum_i  (-1)^{i} (n_i)_{\tilde  m \tilde m} \right] ^2.
\end{eqnarray}
Within the present model, $I$ has been evaluated previously in  Ref.~\onlinecite{mierzejewski16}.

\subsection{Relation to local integrals of motion}

Finite stiffnesses $D_{AA} > 0$ of local operators $A$ are intimately related to the existence of local or quasilocal integrals of 
motion (LIOM),\cite{serbyn13,huse14,serbyn15} and vice versa. 
While this statement has been demonstrated for nearly integrable, translationally-invariant systems,\cite{mierzejewski15} it holds true 
for the present case as well.  For the sake of completeness, we recall here the main arguments.\cite{mierzejewski17} 
Assuming that the summation in Eq.~(\ref{cabst}) is carried  out over all basis states, one finds
 \begin{equation} 
D_{AB} = \frac{1}{N_{tot}} \sum_{\tilde m}  
A_{\tilde  m \tilde m} B_{\tilde  m \tilde m} =\langle  \bar{A} \bar{B} \rangle \label{cabst1},
\end{equation}
where bar means averaging over infinite time-window, $\bar{A}=\sum_{\tilde{m}} A_{\tilde  m \tilde m}  | \tilde m \rangle \langle \tilde m |$. 
Regarding LIOMs, few properties should be emphasized: a) time-averaged operators commute with the Hamiltonian, $[\bar{A},H]=0$;  b) 
the thermal averaging may be considered as the (Hilbert-Schmidt) inner product in the space of local Hermitian operators  
$(A|B)=\langle  A B  \rangle$;  c) time--averaging is an orthogonal projection,   
$\langle  \bar{A} B  \rangle =\langle  A \bar{B}  \rangle =\langle  \bar{A} \bar{B}  \rangle$. 
Consequently, if the stiffness is nonzero,  $D_{AA}=\langle \bar{A}\bar{A}\rangle=||\bar{A}||^2>0$,  
there is a conserved quantity $\bar{A}\ne 0$, which has nonzero projection on strictly local operator $A$,
$D_{AA}=\langle \bar{A} A\rangle$. Therefore, $\bar{A}$ itself is local or quasilocal, hence it must be a LIOM. 
Matrix containing stiffnesses $D_{AB} $ of all local operators $A$ and $B$ allows one to construct 
complete set of orthogonal LIOMs, $Q_l$, for which the so called Mazur bound is 
saturated,\cite{mierzejewski17}
\begin{equation}
D_{AB}  =  \sum_{l} \frac{\langle A Q_l \rangle \langle Q_l B \rangle}{\langle Q_l Q_l \rangle}\;. 
\end{equation} 
Therefore, local or quasilocal operators $Q_l$ fully determine the stiffnesses and vice versa.  
However, the density stiffnesses  $C_d$ are built in terms of only $L$ local operators, $\delta n_1,...,\delta n_L$, and as such 
they  allow one to find at most $L$ local or quasilocal integrals of motion out of
$2^L$ expected for full MBL.\cite{serbyn13,huse14,serbyn15,mierzejewski17}

\subsection{Results}

\begin{figure}[tb]
\includegraphics[width=0.9\columnwidth]{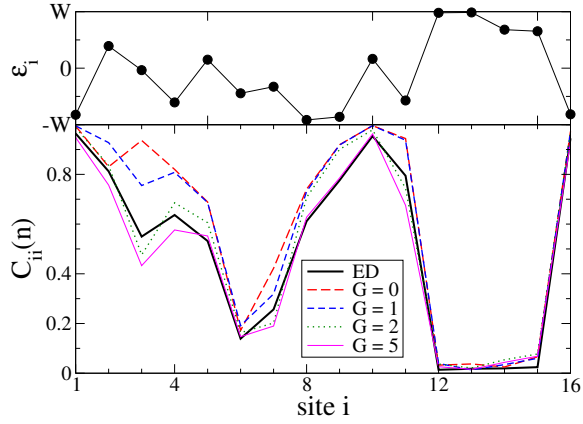}
\caption{(Color online)  Local density stiffness $C_{ii}(n)$, given by Eq.~(\ref{uavstiff}), calculated by RBA for a system of $L=16$ sites, 
$W=6$, $V=1$, and $R=0.1$, while $G$ denotes the number of generations. The upper panel shows the corresponding disorder 
configuration $\epsilon_i$. The RBA results are compared with the exact one.}
\label{fig3}
\end{figure}

With the increasing number of RB states, the RBA results for the local stiffness approach fast the exact results, 
particularly for large disorders, $W \gg W_c$.  Fig.~\ref{fig3} shows the local stiffness obtained without averaging over 
disorder or the initial states $|n\rangle$, cf. Eq. (\ref{cr}),
\begin{equation}
C_{ii}(n)=  \sum_{\tilde m}  |\langle n | \tilde  m \rangle|^2
[(\delta n_i)_{\tilde  m \tilde m}]^2. \label{uavstiff}
\end{equation}
This quantity is obtained from the RBA and is compared with the exact results for $L=16$, $W=6$, $V=1$, 
$R=0.1$ is the resonance threshold parameter used for the RBA, whereas the number of generations is denoted by $G$. In Fig.~\ref{fig3}, 
the persistence of long-time correlations that are strongly site dependent is fully apparent. Furthermore, for the choice of the state 
$|n\rangle$ used to obtain Fig.~\ref{fig3}, some parts of the system exhibit ergodic long-time correlations, $C_{ii}(n)\sim0$, 
whereas other tend to exhibit just the opposite behavior, $C_{ii}(n)\sim1$.

\begin{figure}[tb]
\includegraphics[width=0.9\columnwidth]{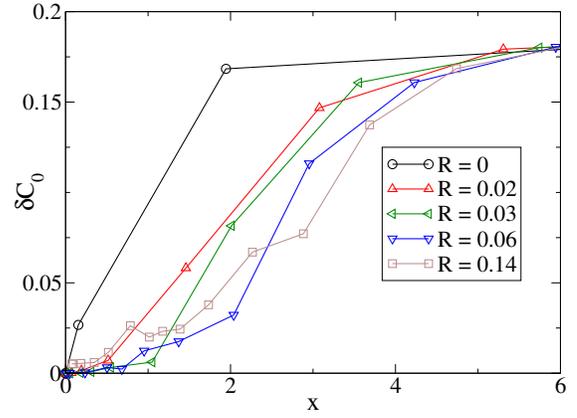}
\caption{(Color online)  An example of deviations of the local density stiffness $\delta C_0$ vs. $x$, Eq.~(\ref{x}), as obtained by the RBA ($L=14$) for an initial state $|n\rangle$ at fixed disorder configuration $\epsilon_i$ corresponding to $W=6$, for different resonance parameters $R$ and increasing number of generations $G$.}
\label{fig4a}
\end{figure}

In order to validate the convergence of RBA, we present in Fig.~\ref{fig4a} a typical example, involving one initial state $|n\rangle$  and one disorder configuration for the system of  $L=14$ sites, with $W=6$. Shown are the results for the standard deviation of local stiffnesses $\delta C_0$, $(\delta C_0)^2= \sum_i (C_{ii}(n) - C^e_{ii}(n))^2/L$, where $C^e_{ii}(n)$ is the exact value given by Eq.~(\ref{uavstiff}). Each curve in Fig.~\ref{fig4a} is obtained for a different value of $R$ by increasing the number of generations $G$. That is, by going from right to left in Fig.~\ref{fig4a}, each symbol along the curves represent a new generation. The results are plotted versus the logarithm of effective number of states, $x$, as defined in
Eq.~(\ref{x}). Since the averaging over configurations and initial states is absent in Fig.~\ref{fig4a}, substantial fluctuations of the $\delta C_0(x)$ curves are observed. Nevertheless, from Fig.~\ref{fig4a} an important conclusion may be drawn. For an optimal choice of $R$ as a function of Hamiltonian parameters, any extreme value of $R$ (e.g., $R=0$ or $R>0.3$) should be avoided. In particular, there is quite a large span
of $R$ values for which the standard deviation $\delta C_0$ scales linearly with $x$. As $W$ is increased, the RBA approaches the exact results at higher values of $x$.

\begin{figure}[tb]
\includegraphics[width=0.9\columnwidth]{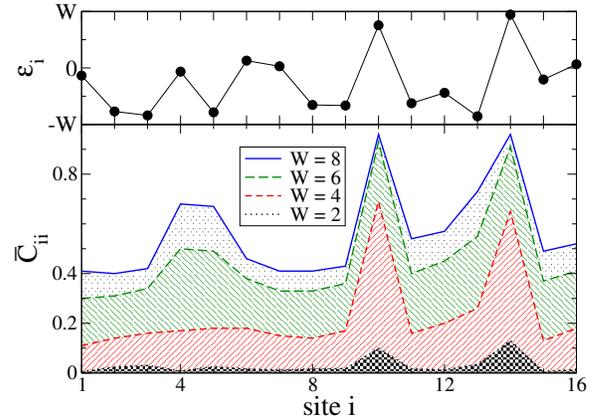}
\caption{(Color online)  Local density stiffness $\bar C_{ii}$, representing the thermal average at fixed configuration of disorder
$\epsilon_i/W$, but for different $W=2-8$,  evaluated for $V=1$ on $L=16$ sites.}
\label{fig4}
\end{figure}

It is plausible that fluctuations of $C_{ii}( n )$ are large in the MBL regime, both among different MB states $| n \rangle$, 
as well as among different configurations of disorder $\epsilon_i$.  However, it is less obvious that in the $T \to \infty$ limit the 
thermal average $\bar C_{ii} = (1/N_{st}) \sum_n C_{ii}(n)$ may depend of the site $i$  as well. Yet, such behavior is clearly 
observed from Fig.~\ref{fig4},  showing the exact results for $\bar C_{ii}$ as a function of the disorder strength $W=2-8$, with $V=1$ 
and with the normalized configurations of disorder $\epsilon_i/W$ held fixed.

In fact, there is a clear correlation between the configurations of disorder $\epsilon_i$, shown in the upper panel of Fig.~\ref{fig4}, 
and the averaged values of $\bar C_{ii}$, shown in the lower panel. In particular, the two most pronounced maxima of 
$\bar C_{ii}$ at $i=10,14,$ correspond to two next-neighbor sites separated by a large barrier, $|\epsilon_i-\epsilon_{i\pm1}|\sim2W$. 
The local values $\bar C_{ii}$ appear to scale with disorder strength $W$.

It is remarkable that $\bar C_{ii}$ in Fig.~\ref{fig4} exhibits such substantial variations between sites,  even for moderate values of the   
disorder strength, $W \geq W_c$. This behavior is consistent with the nonvanishing Edwards-Anderson parameter,\cite{edwards75,luitz116}  
being the hallmark of a glassy (non-ergodic) state. The results in Fig.~\ref{fig4} 
may  be related as well to the concept of rare regions or events that govern the properties of the strongly disordered system.\cite{agarwal15,gopal16,luitz16}

We turn now to results for local stiffnesses $C_0$ averaged over disorder configurations. As noted in Sec.~IIIA, the convenient parameter to follow the convergence is $x$ in Eq.~(\ref{x}). Since, in the RBA context, within the MBL regime both $C_0$ and $N_{st}$ change depending on the initial state $| n\rangle$ and the disorder configuration $\epsilon_i$, the results for $C_0$ are obtained for a fixed $G$ by varying $R$, where $x$ corresponds to the average number of RB states used, $\bar N_{st} =\langle N_{st} \rangle $. One of the major restrictions on accuracy of $C_0$ are statistical fluctuations due to different random configurations $\epsilon_i$, with the relative error decreasing as $\delta C_0 \propto 1/\sqrt{N_r}$.

\begin{figure}[tb]
\includegraphics[width=1.0\columnwidth]{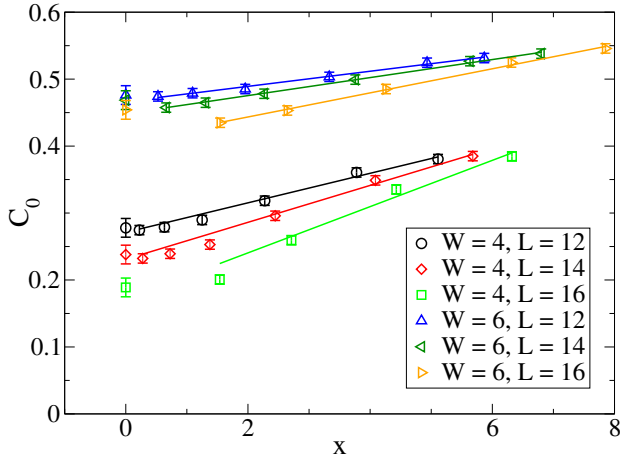}
\caption{(Color online)  Local density stiffness $C_0$ vs. $x$ (logarithm of effective number of states) for
systems $L=12,14,16$ with $V=1$ and $W=4,6$ obtained with RBA, compared at $x=0$ 
with full ED results. Error bars denote standard deviations (statistical error) of calculated $C_0$.}
\label{fig5}
\end{figure}

The RBA is tested in Fig.~\ref{fig5} for system sizes $L \leq 16$ and compared with the ED results at $x=0$. Because of large statistical fluctuations between different disorder samples and initial states, the present numerically most demanding $L=16$ data may suffer a bit from insufficient statistical averaging.  The straight lines in Fig.~\ref{fig5} represents the linear fits of the RBA data for $x>2$.  Few conclusions may be drawn from Fig.~\ref{fig5}: a) The reasonable estimate for $C_0$ is obtained already by using a strongly reduced basis, i.e., for $N_{st} \ll N_{tot}$ corresponding to $x \gg 1$, particularly well in the MBL regime $W \gg W_c$. b) For fixed $L$, most of the points adjust well with the linear scaling of $C_0$ vs. $x$. However, for $x\lesssim2$ a significant deviation from this linear behavior may be observed as well. The reason might be the efficiency of the RBA that increases with $W$, approaching the exact results before $x = 0$ is reached. Whenever RBA  result saturates already for $x>0$, the saturated value  agrees with the corresponding ED result up to statistical errors which arise due to averaging over finite number of disorder configurations.  c) On the MBL side of the transition, the slope of $C_0$ vs. $x$ is only weakly $L$ dependent at given $W$, which allows for sensible extrapolations for larger system sizes $L$.

\begin{figure}[t]
\includegraphics[width=1.0\columnwidth]{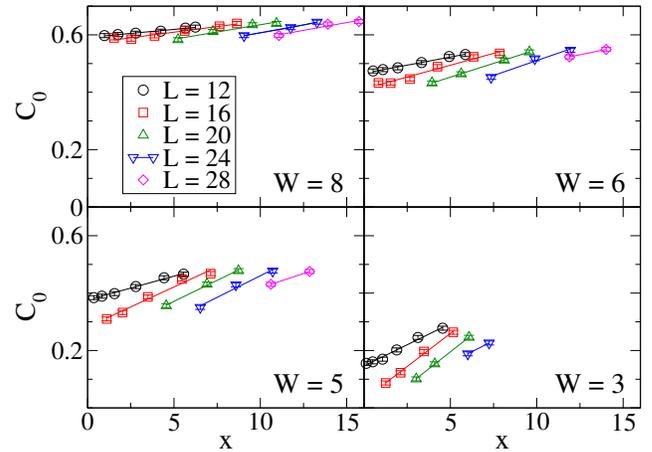}
\caption{(Color online)  Local density stiffness $C_0$ vs. $x$, as obtained by the RBA for systems $L=12 - 28$ and $V=1$, with different $W=3 - 6$. The straight lines 
represent linear fits of the RBA data.}
\label{fig6}
\end{figure}

Representative collection of the RBA results for $C_0$ are presented in Fig.~\ref{fig6} for system sizes up to $L \leq 28$. These results are obtained with fixed $G=6$ and a sampling over disorder configurations $N_c \sim 500-1000$, by lowering $R < 0.8$ until the maximal number of RB states $N_{st}^{max} \sim 15000$ is reached. For large disorders, results reveal evident non-ergodic behavior, e.g., for $W=8 $ the stiffness $C_0$ hardly depends of $x$ at chosen $L$. On the other hand, within the ergodic regime, e.g., at $W=3$, results for $C_0$ exhibit a clear dependence on both $L$ and $x$. The extrapolated behavior can be reconciled with vanishing limiting value $C_0 \to 0$ for $L \to \infty$. 

\begin{figure}[!htb]
\includegraphics[width=1.0\columnwidth]{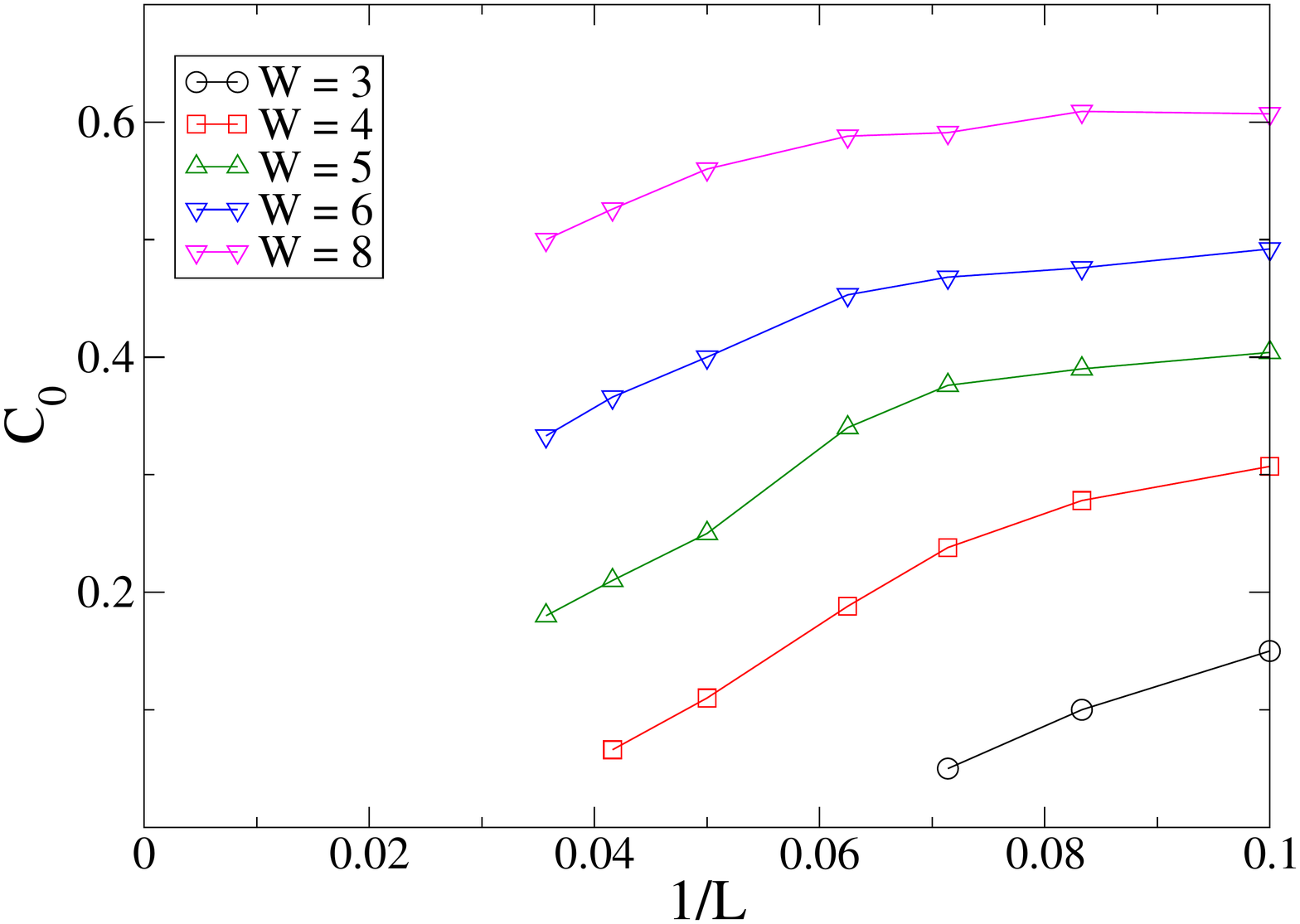}
\caption{(Color online)  $C_0$ as extrapolated to $x=0$ for given $L$ for systems $L=10 - 16$, 
 obtained with the ED, and $L=20 - 28$ with the RBA, corresponding to $V=1$ and different $W=3 - 8$. }
\label{fig7}
\end{figure}

Because of the deviations from the linear behavior of $C_0$ vs. $x$, discussed in connection to Fig.~\ref{fig5}, linear extrapolations of the RBA results towards $x=0$ give the lower limit for the $C_0$ value. The summary of extrapolated $x \to 0$ results for large $L\geq 20$, well beyond the sizes reachable via the full ED, is shown in Fig.~\ref{fig7}. From these results a critical $W_c$ may be estimated, dividing the regimes of limiting $C_0(L \to \infty)=0$ for $W<W_c$ and $C_0(L \to \infty) > 0$ for $W>W_c$. In particular, the results in Fig.~\ref{fig7} for $V=1$ indicate that $W_c \sim 5$, which is a larger value than some previous estimates.\cite{barlev15} However, such $W_c$ may be consistent with other estimations, e.g., $W_c \sim 7.4$ obtained for $V=2$.\cite{luitz15}

\begin{figure}[tb]
\includegraphics[width=1.0\columnwidth]{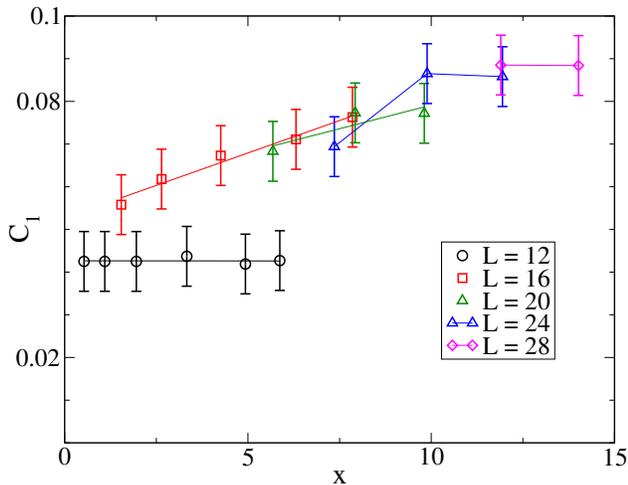}
\caption{(Color online)  Nearest-neighbor density stiffness $C_1$ vs.  $x$ for systems $L=12 - 28$ and fixed 
$V=1$ and $W=6$, obtained with the RBA. }
\label{fig8}
\end{figure}

Results for the nearest-neighbor correlations $C_1$ are presented in Fig.~\ref{fig8} for one selected value of disorder, $W=6$, $W>W_c$. In contrast to $C_0$, the $C_1$ correlations are much weaker, $C_1 \ll 1$, both in the ergodic as well in the MBL regime. In the latter case, $C_1$ vanishes for $W \gg W_c$ due to very short (single-particle) localization length $\xi\ll1$. Nevertheless, as it may be seen from Fig.~\ref{fig8}, in spite of a larger relative (statistical) error, values for $C_1$ are quite well resolved, showing an increasing trend with $L$. The latter behavior may partly be related to the fact that $C_1$ is calculated for a number of particles kept fixed, $N=L/2$, because of which correlations $C_1$ seems to be underestimated when compared to values averaged over grand canonical ensemble.

\section{Survival function}

\subsection{Survival stiffness}

A particularly useful quantity to discuss the validity of the RBA and related approximations is the survival (Loschmidt-echo) function,\cite{heyl14} for a given initial state $|n\rangle$ defined as

\begin{equation}
Q_{n}(t)= |\langle n| \mathrm{e}^{-iHt} | n\rangle|^2\;.
\end{equation}

\noindent Again, we are interested only in the long-time behaviors, i.e., in the survival stiffness $S_n$ and its thermal average $S$ in the $T \to \infty$ limit,

 \begin{equation}
S_n=  \sum_{\tilde m} |\langle n| \tilde m \rangle|^4, \qquad
S =  \frac{1}{N_c} \sum_{n} S_n.\;\label{s}
\end{equation}

\noindent We note that $S$ may be interpreted as the MB inverse participation ratio, which has already been studied in the context of the full ED of the MBL systems.\cite{torres15} Since the distribution of $S_n$ is very broad,\cite{torres15} one may alternatively consider its logarithm, $\ln S_n$. In any case, values of $S$ depend on the initial state $|n \rangle$, as well as on the disorder configurations $\epsilon_i$, because of which the averaging is mandatory for the finite size analysis. Indeed, the physically most interesting is the $L$-dependence of $S$. In contrast to the NI systems, when by construction within the current many-body basis $S=1$, it is generally expected that  $S$ vanishes exponentially with $L$ even in the MBL phase,\cite{heyl14,torres15} 
   
\begin{equation}
 S(L\gg 1) \sim S_0 ~\mathrm{e}^{ - \alpha L}\;,\label{sl}
 \end{equation}

\noindent where $\alpha$ depends on the parameters of the model, in particular on disorder strength $W$. The dependence in Eq.~(\ref{sl}) is given by the number of relevant states $|\tilde m\rangle$. In the case of a MB system satisfying fully the random matrix theory, with the number of particles conserved, one would obtain

\begin{equation}
S(L) \sim N_{tot}^{-1} = \binom{L}{L/2}^{-1}  \sim \mathrm{e}^{-L \ln2}\;,
\end{equation}

\noindent i.e., the maximal value of $\alpha$ is $\alpha_{max} = \ln 2$. It is plausible that for the model under current investigations $\alpha < \alpha_{max}$. In particular,  $\alpha$ should be a decreasing function of $W$. Using $S$, for given $L$ one may as well define the effective number of relevant states, $\tilde N_{st}(L) = 1/S(L)$, where $\tilde N_{st}$ may be used as the criterion whether the RBA is fully or only partially convergent for given $L$ and $W$.

\begin{figure}[tb]
\includegraphics[width=1.0\columnwidth]{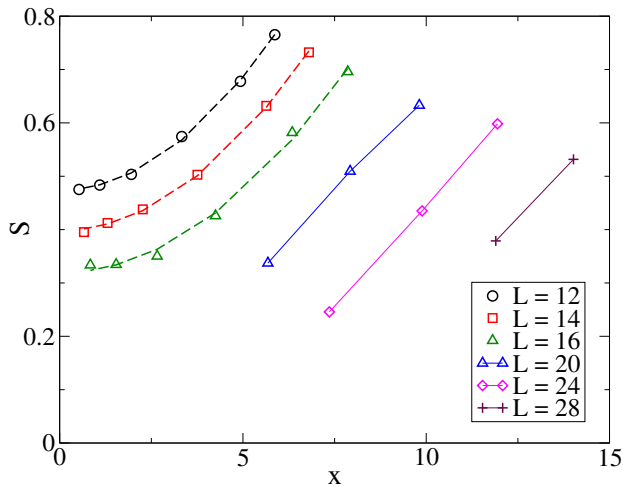}
\caption{(Color online)  Survival stiffness $S$ vs. $x$ for $V=1$, $W=6$, and
various $L=12 - 28$.  }
\label{fig9}
\end{figure}

\subsection{Results}

The survival stiffness $S$, given in Eq.~(\ref{s}), reveals the limitations of the RBA. These limitations arise when one is interested in the precise 
form of the many--body wave function instead of the expectation value of some local observable.
 In Fig.~\ref{fig9}, the RBA results for $S$ vs. $x$ are shown for $W=6 >W_c$, with $S$ averaged over disorder configurations. It is evident from Fig.~\ref{fig9} that $S$ scales in a nontrivial way towards $x=0$. Furthermore, as may be seen from Fig.~\ref{fig9}, for large $L \geq 20$, an extrapolation of results for $S$ towards $x=0$ is not feasible. That is, although we might expect that the RBA involving $N_{st}^{max} \sim 15000$ states should reach expected values down to average $S \ll 0.1$, this is not the case in Fig.~\ref{fig9}. The discrepancy may be explained by particular properties of the survival stiffness, which strongly fluctuates between different disorder configurations. In such circumstances, the average $S$ is dominated by the large values, while the small values of $S$ are already suffering from the RBA limitations.

\begin{figure}[t]
\includegraphics[width=1.0\columnwidth]{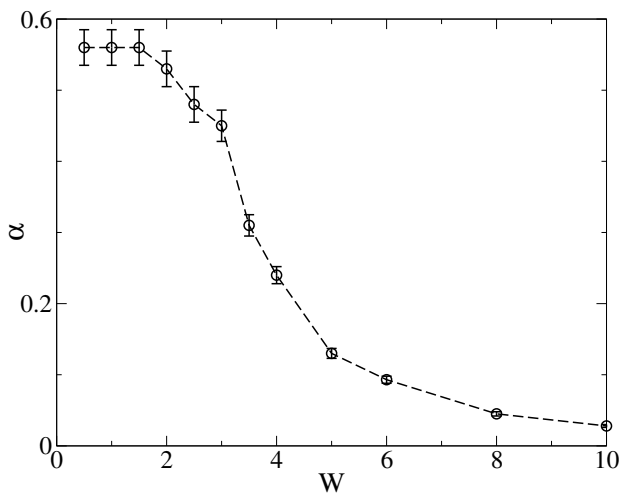}
\caption{Exponent $\alpha$ vs. disorder $W$,
evaluated from the comparison of $L=12$ and $L=16$ ED results.}
\label{fig10}
\end{figure}

Finally, using the exact results for the $L=12$ and the $L=16$ case, in order to estimate $\alpha$ and $S_0$ in Eq.~ (\ref{sl}), we obtain in Fig.~\ref{fig10} the variation of (configuration averaged) exponent $\alpha$ as a function of disorder $W$. It should be emphasized that $S$ depends on the choice of the basis, i.e., our case involves the Anderson MB basis. Nevertheless, the dependence on $W$ in Fig.~\ref{fig10} reveals generic properties of the disordered system. For $W  \ll W_c$, we get a value which approaches (but does not reach) the limiting one $\alpha_{max} =\ln 2 =0.693$, corresponding to the full random-matrix scenario. $\alpha(W)$ reveals the maximal slope in the ergodic regime by approaching the transition $W \leq W_c$. However, we cannot attribute any specific feature of $S(W)$ to the MBL transition. For $W \gg W_c$, we get $\alpha \to 0$ which might be expected for our choice of the MB basis, since for large $W$ corrections to the initial wave function $|n \rangle$ become small, remaining nevertheless $L$-dependent.

\section{Conclusions}

The RBA method presented and its results reveal another perspective regarding the challenging problems of MBL systems. Let us first comment on the feasibility of the method itself. Results of the full ED analysis have the advantage that exact values can be obtained for each disorder configuration up to system sizes $L \leq 16$, where the averaging due to the statistical fluctuations associated with different disorder realizations involve a significant numerical effort for the largest $L=16$. It is evident that the RBA can go beyond the full ED size limitations, in particular if we consider large disorders $W \gg W_c$. We show that well into the MBL regime the quite reliable results can be obtained already with strongly reduced basis $N_{st} \ll N_{tot}$, i.e., $x \gg 1$. On the other hand, closer to the MBL transition, $W \sim W_c$, results (for local density stiffnesses) become more sensitive to changes of $x= \ln(N_{tot}/N_{st})$ and the system size $L$. Starting with some random MB Anderson state, the current numerical implementation of the RBA has two limiting parameters, the resonant criterion $R$ and the number of generations $G$. We did not fully explore the optimization of these two parameters, taking a fixed $G$ throughout most of the calculation and varying $R$ to test the dependence of results on the effective number of RB states $N_{st}$. The way in which the RBA converges towards exact results depends on the parameter regimes. This means that this property alone may be used in different models to locate the transition between ergodic and nonergodic part of the phase diagram, providing additional insights into the structure of wave functions.

This work focuses on long-time limits of correlation functions, i.e., the density stiffnesses $C_d$ and the survival stiffness $S$. These are usually the hardest quantities for numerical evaluations, while at the same time being the essential hallmarks of the MBL. In the thermodynamic limit $L \to \infty$, within the MBL regime, the averaged density stiffnesses $C_d$ should remain finite for $d=0$, as well as for small $d>0$, due to their overlaps with the LIOMs. Namely, the latter are (quasi)local operators as well and represent an alternative defining signature of the MBL. Our analysis shows that for large systems above the reach of the full ED, $C_d$ are well represented even within very restricted $N_{st} \ll N_{tot}$ bases. In order to get more accurate results for $C_d$ an extrapolation to $N_{st} \sim N_{tot}$ is needed and we show that the convenient extrapolation parameter is $\ln N_{st}$, or $x$ given by Eq.~(\ref{x}). Such dependence of $x$ has the origin in the marginal validity of the perturbation expansion (in $V$) of local quantities as $C_d$ at fixed $L$. Moreover, it appears that such property is quite generic in strongly correlated systems.\cite{prelovsek93}

The limitations of the RBA are more apparent in the context of the survival stiffness $S$, since it is impossible for the RBA to reproduce results with very small $S < 1/N_{st}$. That is, the eigenfunctions of the interacting disordered systems cannot be fully described by the RBA for $S \ll 1$, which is clearly the case in the ergodic regime $W< W_c$, as well as in the nonergodic regime at large $L > L^*(W) \sim 1/\alpha(W)$. In relation to $S$, the most relevant parameter is the survival exponent $\alpha$. As a function of $W$, $\alpha$ has the strongest variation below the transition $W < W_c$, remaining finite but small ($\alpha \ll 1$) in the MBL regime $W> W_c$.

The finite local density stiffness $C_0>0$ is the most obvious hallmark of the MBL regime. The results for a single initial state $C_{ii}(n)$ (as in Fig.~\ref{fig3}) indicate that for $W>W_c$ this quantity exhibits large fluctuations among different sites $i$. Remarkably, in the MBL regime, some of the sites or regions exist for which $C_{ii}(n)\sim0$, impling nearly the ergodic local behavior. This effect remains well visible even after full thermal averaging for a fixed disorder configuration $\epsilon_i$, as shown in Fig.~\ref{fig4}. It appears, as a general property, that close to the transition the thermal $ \bar C_{ii}$ are dominated by large fluctuations, developing above a quite uniform and even small $C_0$. This emphasizes the importance of rare regions, frequently invoked in the interpretations of the MBL,\cite{agarwal15,gopal15,luitz16} both from the ergodic and non-ergodic side.

The results for the averaged $C_0$ at fixed $L$ generally agree with previous ED findings for $L \leq 16$. The extrapolated $x \to 0$ results obtained by the RBA confirm further decrease of $C_0$ with $L$. As a consequence, for $V=1$ previous estimates for the transition\cite{barlev15,mierzejewski16} seem to underestimate the critical $W_c$, i.e., we obtain $W_c \sim 5$. Our estimate seems to be more consistent with the estimations based on the level statistics, e.g., $W_c \sim 7.4$ for $V=2$.\cite{luitz15,luitz16} On the other hand, extrapolations beyond $L \gg 20$ require some caution as well. Therefore, the current predictions for the limit $L \to \infty$ are not fully conclusive in this respect. 

Since the presented RBA method can deliver quite reliable results in the MBL regime even for strongly reduced
basis $N_{st} \ll N_{tot}$, it may be applied well beyond the standard 1D MBL model, as tested in the 
present work. It is plausible that this approach can be used for the analysis of higher-dimensional disordered MB models, which is quite
unexplored area so far, as well as for more complex 1D models, in particular the Hubbard model realized in
optical lattice experiments. Moreover, one may extend its application to time-dependent correlations, which should
further elucidate the extremely slow relaxation in the MBL systems.

\begin{acknowledgments}

P.P. acknowledges the support by the program P1-0044 of the Slovenian Research Agency.  
O.S.B. acknowledges the support by Croatian Science Foundation Project No. IP-2016-06-7258. 
M.M.  acknowledges support by the National Science Centre, Poland via project 2016/23/B/ST3/00647.

\end{acknowledgments}


%

\end{document}